\documentclass[10pt,prl,aps,twocolumn,superscriptaddress,floatfix,showpacs]{revtex4}

\usepackage{graphicx}
\usepackage{amssymb,amsmath}
\usepackage{epstopdf}
\usepackage{bm}
\usepackage{color}

\begin{document}

\title{Numerical detection of symmetry enriched topological phases with space group symmetry}

\author{Ling Wang}
\affiliation{Institute for Quantum Information and Matter, California Institute of Technology, Pasadena, California 91125, USA}
\affiliation{Department of Physics, California Institute of Technology, Pasadena, California 91125, USA}
\author{Andrew Essin}
\affiliation{Institute for Quantum Information and Matter, California Institute of Technology, Pasadena, California 91125, USA}
\affiliation{Department of Physics, California Institute of Technology, Pasadena, California 91125, USA}
\author{Michael Hermele}
\affiliation{Department of Physics, 390 UCB, University of Colorado, Boulder CO 80309, USA}
\author{Olexei Motrunich}
\affiliation{Department of Physics, California Institute of Technology, Pasadena, California 91125, USA}

\date{\today}

\begin{abstract}
  Topologically ordered phases of matter, in particular so-called
  symmetry enriched topological (SET) phases, can exhibit quantum
  number fractionalization in the presence of global symmetry.  In
  $\mathbb{Z}_2$ topologically ordered states in two dimensions,
  fundamental translations $T_x$ and $T_y$ acting on anyons can either
  commute or anticommute. This property, crystal momentum
  fractionalization, can be seen in a periodicity of the excited-state
  spectrum in the Brillouin zone.  We present a numerical method to
  detect the presence of this form of symmetry enrichment given a
  projected entangled pair state (PEPS); we study the minima of
  spectrum of correlation lengths of the transfer matrix for a
  cylinder. As a benchmark, we demonstrate our method using a modified
  toric code model with perturbation. An enhanced periodicity in
  momentum clearly reveals the nontrivial anticommutation relation
  $\{T_x,T_y\}=0$ for the corresponding quasiparticles in the system.
\end{abstract}

\maketitle

Topological order is the name given to a variety of long range
entangled but gapped phases of matter, in particular to phases that
support anyonic excitations with unusual braiding statistics. Unlike
phases that fall within the Landau symmetry-breaking paradigm,
topological order is defined without any reference to symmetry.
However, it is still very interesting to ask what further phenomena
emerge in systems with symmetry, either spontaneously broken or
unbroken. The term ``symmetry enriched topological'' (SET) phases was
proposed to describe phases that have the same topological order but
are distinct in the presence of a
symmetry~\cite{Hung2013,andrew2,yuanminglu}.

In two dimensions the excitations in a topological phase are
point-like anyons. When a symmetry is present, what are their quantum
numbers?  It turns out that these can be fractional; most prominently,
anyons in quantum Hall states typically have fractional electric
charge (the quantum number corresponding to a U(1) symmetry of the
system)~\cite{tsui,laughlin}. The values of these quantum numbers are
highly constrained, notably by the ``fusion rules'' of the topological
theory; in the Laughlin quantum Hall state with filling fraction
$1/3$, since three anyons give back an electron, the anyons must have
charge $e/3$. Another example is the spin fractionalization in spin
liquid phases of quantum magnets with a global $\text{SU}(2)$
symmetry: a spinon quasiparticle carries a fractional number
spin-$1/2$ whereas in conventional paramagnets or magnetically ordered
states all quasiparticles carry integer spin~\cite{balents}.

Can one distinguish SET phases given a wavefunction? The quantum
numbers of the degenerate ground states and the projective quantum
numbers of quasiparticles are the characteristic properties of the SET
phases~\cite{jalabert,paramekanti,kou,andrew2}. In the case of
detecting the projective quantum number of an internal symmetry, the
operation of the global internal symmetry generator can be factorized
into a product of local operators, which can be transformed into
operators acting only on the entanglement cut of the system, therefore
one can detect the projective quantum number via measuring the
commutator/anti-commutator of the boundary
quasiparticles~\cite{huang,pollmann1}. However for the space group
symmetry, such as the translation symmetries, the translation operator
is written in a matrix product operator that can not be factorized,
and in addition, there is no way to make an entanglement cut that
preserves both $T_x$ and $T_y$ translations, therefore conventional
techniques do not work. In this Letter, we address this question for
space group symmetries in the context of projected entangled pair
states (PEPSs)~\cite{frank04}.

{\it $\mathbb{Z}_2$ topological phases with translation symmetry} --
We are interested in the topological order familiar from
$\mathbb{Z}_2$ gauge theory, $\mathbb{Z}_2$ spin liquids, and the
toric code.  There are two bosonic anyon species, often called $e$ and
$m$.  Each sees the other with an Aharonov-Bohm phase of $-1$ (they
are mutual semions).  When two-dimensional translation symmetry is
present, the symmetry generators $T_x$ and $T_y$ may act nontrivially
(projectively) on the anyons~\cite{andrew2,andrew1}. A
basis-independent characterization of these actions is given by the
relations
\begin{eqnarray} \label{eq:rels}
\nonumber
T_x^e T_y^e T_x^{e-1} T_y^{e-1} &=& \eta_e = \pm 1, \\
T_x^m T_y^m T_x^{m-1} T_y^{m-1} &=& \eta_m = \pm 1,
\end{eqnarray}
where $T_x^e$ is the action of $T_x$ on a single $e$, etc.  When one
of these relations evaluates to $-1$ we say that translations act
projectively, or that the anyon has fractional crystal momentum or has
nontrivial fractionalization class, a notion closely related to Wen's
projective symmetry group~\cite{wen_psg}.

One can interpret the nontrivial $e$ relation as the presence of an
$m$ in each unit cell of the lattice, which the $e$ sees as a
background $\pi$ magnetic flux.  It is straightforward to show, based
on this observation or directly from the relations above, that if $e$
has fractional crystal momentum, the spectrum (and density of states)
of two-$e$ scattering states is periodic under $\mathbf{q} \rightarrow
\mathbf{q} + (\pi,0), (0,\pi),
(\pi,\pi)$~\cite{Wen2002,andrew1}. Assuming that $e$ is the excitation
of lowest energy, the low-energy edge of the continuum of excited
states will reveal the fractionalization class of $e$ in the dynamical
structure factor of any operator that excites anyons. This is the main
idea for detecting such SET states that we pursue in this
paper. However, we access the low-energy edge of the excited states
via the information contained in the ground state instead of the
excitation spectrum due to a nice property given by the PEPS, which we
will introduce below.

{\it PEPS and transfer matrix --}
PEPS is an ansatz that represents the wavefunction by locally
entangled virtual pairs and a projector that map the virtual system
to the physical one~\cite{frank_pepsreview}.  It captures a wide
range of phases including many with topological
order~\cite{frank06,gu_terg,gu_stringnet,vidal_stringnet,cirac_ES,didier_rvb,norbert_rvb,norbert2d}.
A one-dimensional version, the MPS~\cite{ostlundrommer}, has been used
to classify phases of symmetric, gapped spin systems based on the
projective representation of the symmetry
group~\cite{xie1d,norbert1d,Pollmann2010}. Here, we propose a method based on PEPS
for the spectra of correlation lengths (SCL) of the system, which
allows us to distinguish SET phases described by simple PEPSs.

\begin{figure}
\begin{center}
\includegraphics[width=7.5cm]{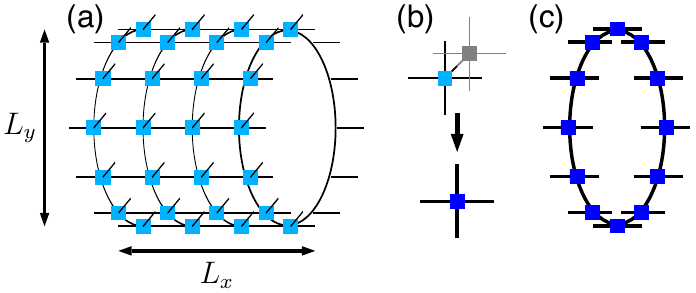}
\caption{(a) Schematic representation of a PEPS on a cylinder of size
  $L_x\times L_y$. (b) When multiplying bra and ket of PEPS, we sum
  over the physical degrees of freedom, group the virtual indices of a
  bond, and arrive at the double tensor of a single site. (c) Placing
  $L_y$ double tensors on a ring and tracing out the virtual degrees
  on the shared bonds, we form the transfer matrix of the cylinder.}
\label{cylinder}
\end{center}
\vskip -0.5cm
\end{figure}

In one-dimensional gapped systems, properties of the ground state wave
function (the MPS) are completely determined by the transfer matrix if
translation symmetry is present. The spectrum of correlation lengths,
which is given by the negative of the logarithm of (normalized)
eigenvalues ($\lambda$) of the transfer matrix~\cite{ostlundrommer},
is intuitively related to the spectrum of excitations made by all
possible local operators with momentum quantum number $k_x$,
$\lambda=|\lambda|e^{ik_x}$, at the minima of the
spectrum~\cite{zauner}.

The SCL can be generalized in two dimensions, however the momentum
quantum number in the $y$ direction $k_y$ is rigorously well defined
as compared to the $k_x$ estimated from the complex phases of the
transfer matrix eigenvalues, as in the one-dimensional case mentioned
above. Consider a cylindrical geometry (as in Fig.~\ref{cylinder}), we
can define the SCL of the transfer matrix of a cylinder. If
translation symmetry in the $y$ direction is present, eigenvectors of
the transfer matrix have well defined momentum quantum numbers $k_y$,
and the minimum of the SCL is given by
\begin{equation}
  \epsilon_{(k_x,k_y)}=-\text{ln}(|\lambda_{(k_x,k_y)}^{\text{max}}|/\lambda_0),
\end{equation}
where
$\lambda_{(k_x,k_y)}^{\text{max}}=e^{ik_x}|\lambda_{(k_x,k_y)}^{\text{max}}|$
is the leading eigenvalue of the transfer matrix with momentum $k_y$
excluding ground states $\lambda_0$s (the largest eigenvalue among all
sectors); in general it is a complex number with a phase $e^{ik_x}$
where $k_x$ is its momentum in the $x$ direction~\cite{zauner}. We
conjecture that $\epsilon_{(k_x,k_y)}$ at a given $(k_x,k_y)$ are
analogous to the low-energy edge of the two-anyon scattering continuum
described earlier.  Thus, we propose that the minima of the SCL can be
used to distinguish SET phases, by analogy with the dynamic structure
factor. Next we will test this conjecture by examining the minima of
the SCL of the transfer matrix.

\begin{figure}
\begin{center}
\includegraphics[width=6cm]{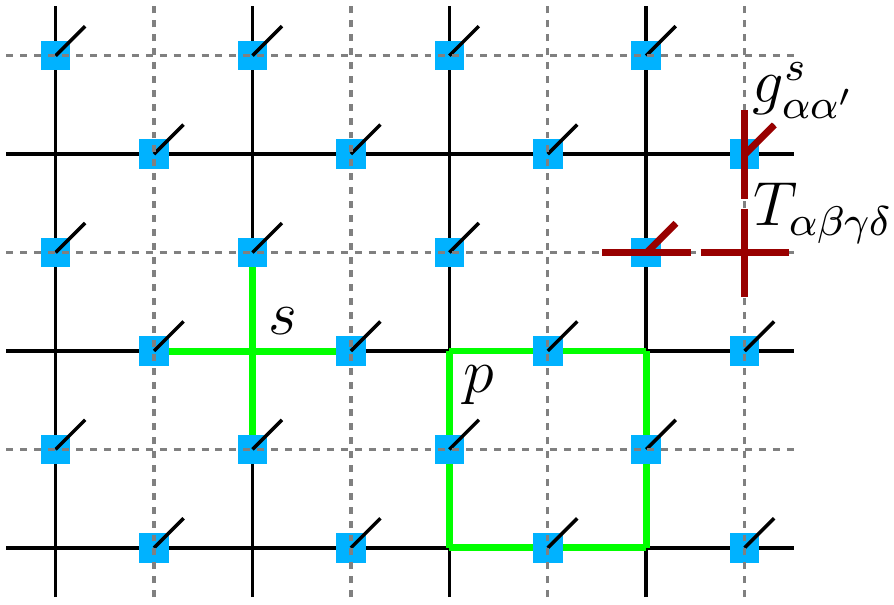}
\caption{A demonstration of the PEPS wavefunction of the toric code
  model in the $\sigma^z$ basis. Green lines denote the star operator
  $A_s$ and the plaquette operator $B_p$, and dark red lines denote
  the position of the 4-index $T$ tensor and the 3-index $g$ tensor,
  which are placed at the vertices and bonds of the dual lattice
  (denoted in dashed lines).}
\label{toriccode_peps}
\end{center}
\end{figure}

{\it Modified toric code model} -- To construct the simplest
$\mathbb{Z}_2$ spin liquid which realizes all possible projective
quantum numbers of the translation symmetry, we consider the toric
code Hamiltonian on a square lattice
\begin{equation}
\label{toriccode_hamiltonian}
H=-K_e\sum_sA_s-K_m\sum_pB_p,
\end{equation}
where $A_s=\prod_{l\in s} \sigma_{l}^x$ is defined on the vertex $s$ and
$B_p=\prod_{l\in p}\sigma_{l}^z$ is defined on the plaquette $p$, and
the sum runs over all vertices and all plaquettes. This Hamiltonian
has four degenerate ground states~\cite{kitaev}. Depending on the signs
of $K_m$ and $K_e$, the quasi-particles $e$, $m$, if created, will move
in a background of $0$- or $\pi$-flux.  That is, $\eta_e = \mathrm{sign} K_m$, $\eta_m = \mathrm{sign} K_e$ in Eq.~\eqref{eq:rels}.  The ground states on a torus
can be simply represented by PEPSs of bond dimension $D=2$. We now
describe the PEPSs for all choices of $K_e=\pm 1$ and $K_m=\pm 1$.

The PEPSs are composed of 4-index tensors
$T_{\alpha\beta\gamma\delta}$ at the vertices and 3-index tensors
$g^s_{\alpha\alpha^{\prime}}$ at the bonds of the direct (or dual)
lattice, where $s$ represents physical degrees of freedom and Greek
letters represent virtual ones. Whether we take the direct or dual
lattice depends on the choice of using a local $\sigma^x$ or
$\sigma^z$ basis.  Figure~\ref{toriccode_peps} represents a PEPS
defined in the $\sigma^z$ basis, where the tensor
$T_{\alpha\beta\gamma\delta}$ is placed at the vertices of the dual
lattice; this is the representation we choose throughout this paper
unless specified otherwise. The virtual index runs from $0$ to $1$,
where 0 means $|\uparrow\rangle$ and 1 means $|\downarrow\rangle$. The
wave function has the form $|\psi\rangle=\text{Tr}\{T^{\otimes
  V}g^{\otimes B}\}$, where $V$ means all vertices and $B$ means all
bonds, and the trace is taken over all common virtual degrees. The
$T$-tensor is
\begin{equation}
\label{ttensor}
T_{\alpha\beta\gamma\delta}=
\begin{cases}
1,(0,) & (\alpha+\beta+\gamma+\delta)\%2=0 \\
0,(1,) & (\alpha+\beta+\gamma+\delta)\%2=1 
\end{cases},
\end{equation}
which, together with the condition that the only non-zero elements of
the $g$-tensor are $g^{s}_{ss}$ (see below), enforces the condition
$B_p|\psi\rangle = + (-)|\psi\rangle$, corresponding to $K_m>0$
($K_m<0$).  The elements of the $g$-tensor are
\begin{equation}
\label{stensor}
g^s_{\alpha\alpha^{\prime}}=
\begin{cases}
a, & \alpha=\alpha^{\prime}=s=0\\
1, & \alpha=\alpha^{\prime}=s=1\\
0, & \text{otherwise},
\end{cases}
\end{equation}
where $a$ is some number which can also depend on the position of the
bond.  We ask that the wave function satisfies
$A_s|\psi\rangle=+(-)|\psi\rangle$ for $K_e>0$ ($K_e<0$); this is
nothing but asking that the amplitudes for configurations related by
flipping the four spins on the vertex $s$ differ by $+$ ($-$). If
$K_e>0$ the solution is obvious: $a=1$ on all bonds. However, if
$K_e<0$, it requires one and only one $a$ on each plaquette of the
dual lattice to be -1, in which case one has to break the lattice
translation symmetry in one direction in order to keep the bond
dimension $D=2$ (Supplemental Material~\cite{supp}).

The excitations of the Hamiltonian Eq.~(\ref{toriccode_hamiltonian})
have no dynamics, which corresponds in the ground state $|\psi\rangle$
to the fact that all spin configurations satisfying
Eq.~(\ref{ttensor}-\ref{stensor}) have equal weight in magnitude. To
create some dynamics without increasing the bond dimension of the
tensor, we put a local diagonal operator $\text{diag}(w,1)$ on each
physical spin with $0\le w \le 1$, which corresponds to a ground state
of the Hamiltonian~\cite{Castelnovo,Schuch2013}
\begin{align}
\label{toriccode_hamiltonian2}
H' &= H + K_e \sum_s w^{-\sum_{l\in s} \sigma^z_l} \notag\\
&\approx H + h \sum_l \sigma^z_l + \text{const}.
\end{align}
Here, the Zeeman field $h = -2K_e \log w$ is a good description of the
perturbed state for $h \ll K_e$, or $w \approx 1$.  Throughout this
paper, we use $w=0.9$, which corresponds to $h\ll |K_m|,|K_e|$ and is
deep in the topologically ordered phase. For the phase diagram of the
modified PEPS as a function of $w$ in case of $K_m<0$ and $K_e>0$
(Supplemental Material~\cite{supp}) .

\begin{figure}
\begin{center}
\includegraphics[width=7.5cm]{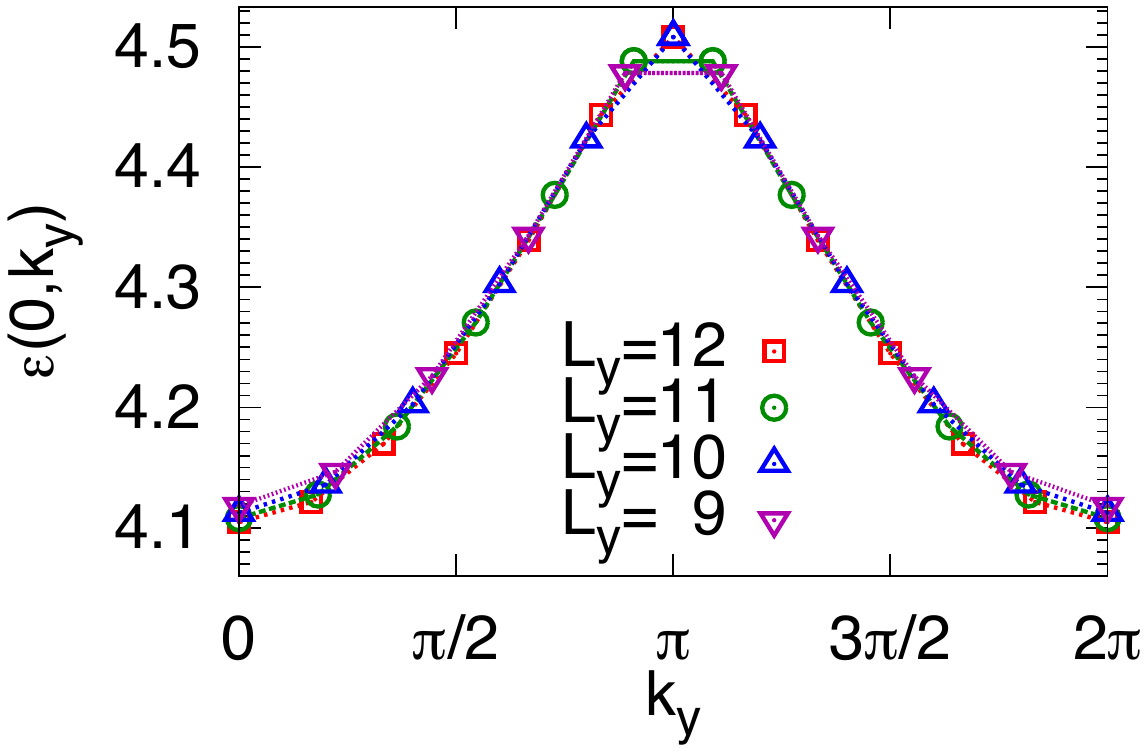}
\caption{The minima of SCL at momentum $k_y$ for coupling $K_m>0$ and
  $K_e>0$ using PEPS at $w=0.9$. Since all eigenvalues of the transfer
  matrix are real and positive, only $\epsilon_{(0,k_y)}$ is
  presented.}
\label{scl2}
\end{center}
\vskip -0.5cm
\end{figure}

\begin{figure}
\begin{center}
\includegraphics[width=7.4cm]{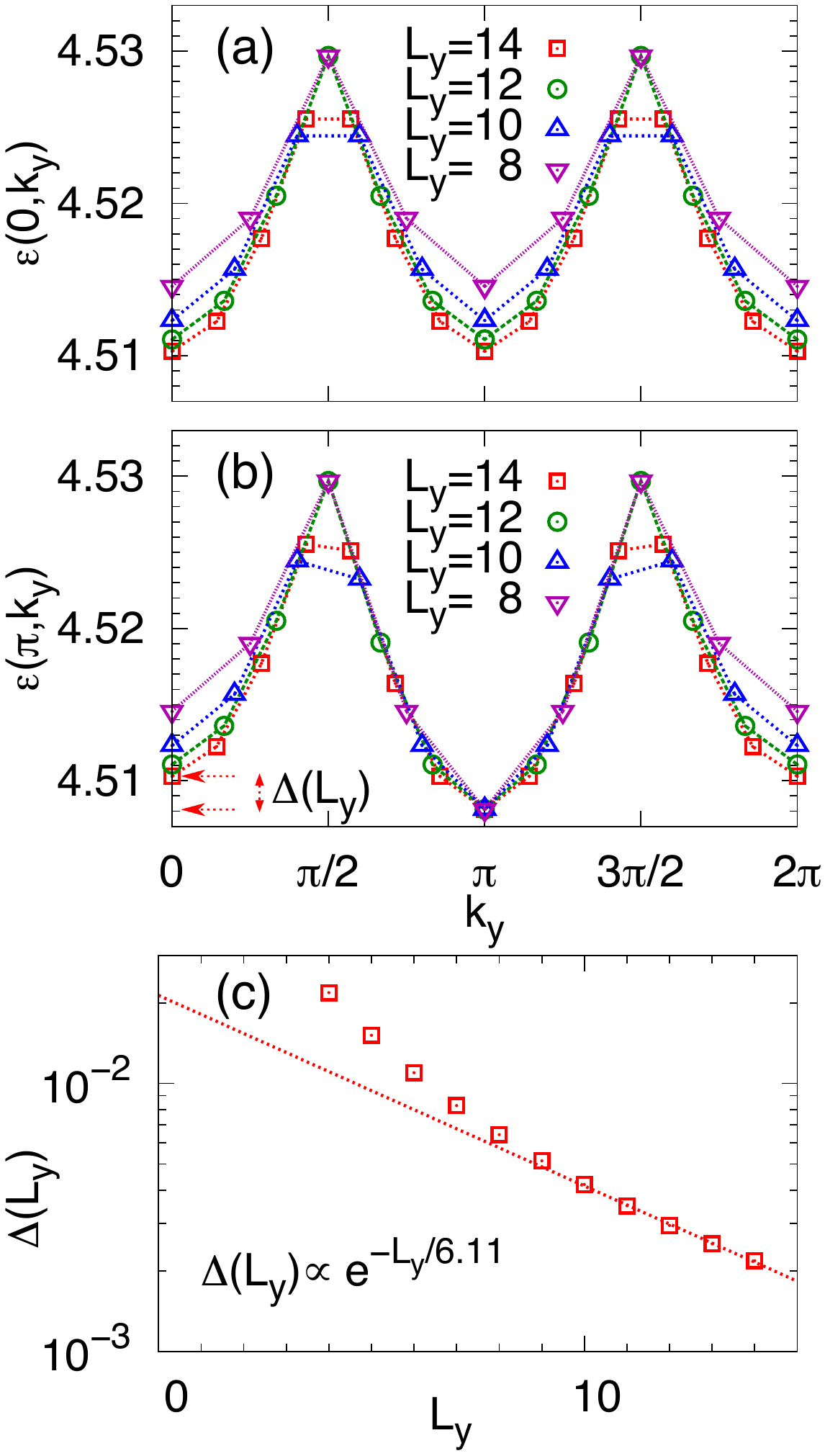}
\caption{The minima of SCL at momentum $k_y$ for coupling $K_m<0$ and
  $K_e>0$ using PEPS at $w=0.9$ for even $L_y$. (a)
  $\epsilon_{(0,k_y)}$ corresponds to the largest real and positive
  $\lambda$ in sector $k_y$ except $\lambda_0$s. (b)
  $\epsilon_{(\pi,k_y)}$ corresponds to the smallest real and negative
  $\lambda$ in sector $k_y$. (c) The splitting
  $\Delta(L_y)=\epsilon_{(\pi,0)}-\epsilon_{(\pi,\pi)}$ as illustrated
  in (b) vanish exponentially as a function of length $L_y$.}
\label{scl1}
\end{center}
\vskip -0.5cm
\end{figure}

\begin{figure}
\begin{center}
\includegraphics[width=7.4cm]{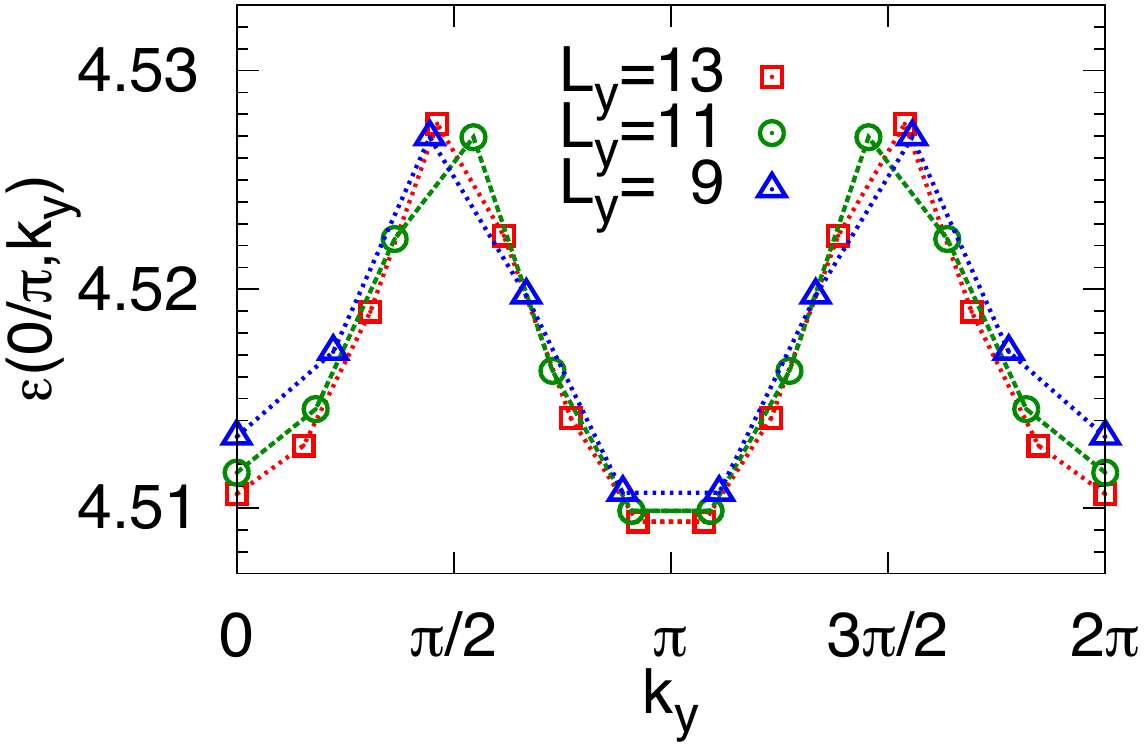}
\caption{The minima of SCL at momentum $k_y$ for coupling $K_m<0$ and
  $K_e>0$ using PEPS at $w=0.9$ for odd length $L_y$. Since
  eigenvalues including the largest come in $\pm$ pairs,
  $\epsilon_{(0,k_y)}$ and $\epsilon_{(\pi,k_y)}$ are identical.}
\label{scl3}
\end{center}
\vskip -0.5cm
\end{figure}

{\it Numerical detection of the projective quantum number under
  translation symmetries} -- We take the PEPS wave functions described
above and calculate the SCL of the transfer matrix [see
Fig.~\ref{cylinder}(b)] for different signs of $K_m$ and $K_e$. The
transfer matrix of a cylinder with translation symmetry $T_y$ can be
block-diagonalized in momentum basis, which means each left and right
eigenvector has a well defined momentum quantum number $k_y$. We start
by explicitly writing the transfer matrix in real space into a block
diagonal form in momentum space (Supplemental
Material~\cite{supp}). Once we have obtained the transfer matrix in
momentum basis, we diagonalize each block with momentum $k_y=2\pi
m/L_y, m=0,1,\cdots,L_y-1$, find the minima of the normalized SCL
$\epsilon_{(k_x,k_y)}=-\text{ln}(|\lambda_{(k_x,k_y)}^{\text{max}}|/\lambda_{0})$,
and plot $\epsilon_{(k_x,k_y)}$ as a function of $k_y$ for $k_x=0$ and
$\pi$. Fig.~\ref{scl2} illustrates the minima of the SCL for $K_m>0$
and $K_e>0$, in which case, all eigenvalues of the transfer matrix are
real and positive. The results for $K_m<0$, $K_e>0$ and $L_y$ even are
presented in Fig.~\ref{scl1}: $\epsilon_{(0,k_y)}$ corresponds to the
largest real and positive eigenvalue $\lambda$ at each $k_y$ excluding
the two degenerate $\lambda_0$s at $k_y=0$, while
$\epsilon_{(\pi,k_y)}$ corresponds to the smallest real and negative
eigenvalue $\lambda$ at each $k_y$. In the case of odd $L_y$ for $K_m
< 0$ and $K_e > 0$, matrices at each $k_y$ further form into two
non-commuting blocks, encoding the fact that applying the transfer
matrix flips the eigenvalue of $\prod_l \sigma^z_l$, where the product
is over a loop encircling the system in the $y$-direction.  This
corresponds to a breaking of translation symmetry ($T_x$) in a
one-dimensional limit ($L_x \to \infty$ with $L_y$ fixed).  All
eigenvalues of the transfer matrix come in $\pm$ parirs, thus
$\epsilon_{(0/\pi,k_y)}$ are identical, as presented in
Fig.~\ref{scl3}. We find that the minima of the SCL are indeed doubly
periodic when $K_m<0$, but not when $K_m>0$.

For the case of $K_e<0$ and $K_m<0$ ($K_m>0$), the resulting
$\epsilon_{(0,k_y)}$ are exactly the same as above, because the
low-energy excitations are dictated by coupling $K_m$ regardless of
the sign of $K_e$. Note that when $K_e<0$, $\epsilon_{(\pi,k_y)}$ is
not accessible, because the transfer matrix at any $L_y$ consists of
adjacent two columns of the lattice, thus all eigenvalues are
positive.

If we want the projective quantum number of $m$-particles, we can take
a perturbed Hamiltonian as
\begin{equation}
\label{toriccode_hamiltonian3}
H''=-K_e\sum_sA_s-K_m\sum_pB_p + h'\sum_l\sigma_l^x
\end{equation} 
instead, and correspondingly choose a PEPS representation defined in
the $\sigma^x$ basis, where acting with a local operator
$\text{diag}(w,1)$ is equivalent to adding a perturbation $\delta H =
h'\sum_l\sigma^x_l$. The low-energy excitations are two $m$-particles
and the sign of $K_e$ determines whether the $m$-particle hops in a
background of 0- or $\pi$-flux on the dual lattice. Once the
projective quantum numbers $\eta_e$ and $\eta_m$ are known, one
completely determines SET class of the state if the symmetry group
consists only of translation~\cite{andrew2}.

{\it Conclusion} -- We have presented a numerical method to
distinguish symmetry enriched topological (SET) phases with
fractionalized translation symmetry. This method uses the spectrum of
correlation lengths (SCL) of the transfer matrix on a cylinder to
represent qualitatively the excitation spectrum of the system as a
function of $k_y$ for special $k_x=0,\pi$, if the ground state wave
function is available in terms of projected entangled pair state
(PEPS). From the fact that the nontrivial fractional quantum numbers
of quasi-particles under translation symmetries will be manifest as
enhanced Brillouin zone periodicity in the dispersion relation, one
can read out the projective quantum number of the low energy
quasiparticles from the behavior of the minima of SCL. We bench-marked
this method with the toric code model under perturbation. Modifying
the sign of the coupling coefficients in front of operator $A_s$ and
$B_p$ and the perturbation terms, we were able to generate
topologically ordered ground states with preferred $e$-particle or
$m$-particle low-energy excitations in a background of either $0$ or
$\pi$ magnetic flux, which realizes all symmetry classes with this
topological order and symmetry~\cite{andrew2}.  We expressed the
ground states of the modified toric code Hamiltonians as
bond-dimension $D=2$ PEPSs and calculated the minima of the SCL of the
transfer matrix on a cylinder, and the pattern of SCL revealed the
fractional quantum numbers of the low energy quasiparticles under
translation symmetries.

This method can be generalized to detect projective quantum numbers of
SET phases under a broader symmetry group including translations and
other space group symmetries.

{\it Acknowledgment} -- We would like to thank F.~Verstraete and
R.~Mong for useful discussion.  This work was supported by the
Institute for Quantum Information and Matter, an NSF Physics Frontiers
Center with support of the Gordon and Betty Moore Foundation through
Grant GBMF1250, by the U.S. Department of Energy (DOE), Office of Science,
Basic Energy Sciences (BES) under Award \# DE-FG02-10ER46686 (M.H.),
by Simons Foundation grant \# 305008 (M.H. sabbatical support), and
by the National Science Foundation through grant DMR-1206096 (O.M.).

\section{Appendix}

\subsection{PEPS for the ground state of toric code model with $K_e<0$}
In this section, we will discuss how to write a PEPS ground state for
the toric code model with $K_e<0$ while keeping the bond dimension
$D=2$.

\begin{figure}
\begin{center}
\includegraphics[width=7.5cm]{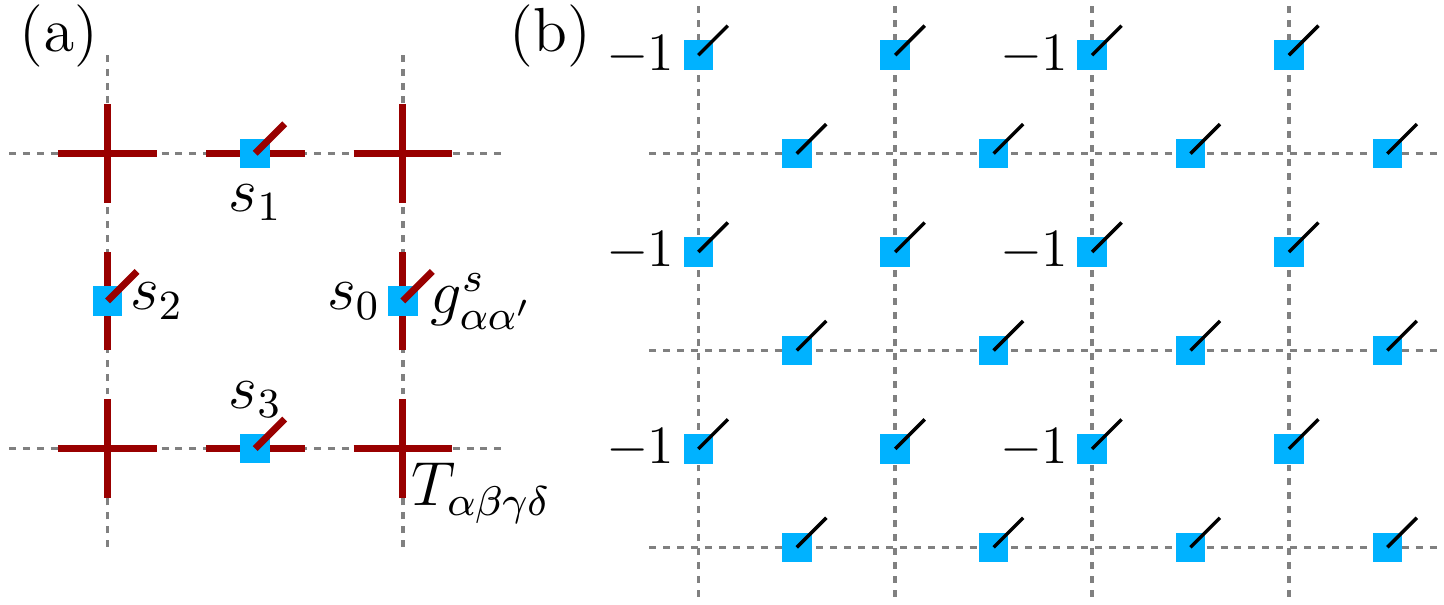}
\caption{(a) Demonstrate the vertices $T_{\alpha\beta\gamma\delta}$
  and bonds $g_{\alpha\alpha^{\prime}}^s$ affected by the operation of
  $A_s$ on one plaquette of the dual lattice. (b) To meet the
  condition $A_s|\psi\rangle=-|\psi\rangle$, we choose $a=-1$ on the
  vertical bonds of every other column of the dual lattice (as marked
  by $-1$) and $a=1$ for all other bonds (without mark).}
\label{ke_negative}
\end{center}
\end{figure}

We want to write the PEPS in the $\sigma_z$ basis since our
perturbation $h\sum_l\sigma_l^z$ is simple in this
basis.  The PEPS is composed of a 4-index tensor
$T_{\alpha\beta\gamma\delta}$ of bond dimension $D=2$
\begin{equation}
\label{ttensor1}
T_{\alpha\beta\gamma\delta}=
\begin{cases}
1,  (0,)  & (\alpha+\beta+\gamma+\delta)\%2=0 \\
0,  (1,)  & (\alpha+\beta+\gamma+\delta)\%2=1 ,
\end{cases}
\end{equation}
for $K_m>0$ ($K_m<0$), and a 3-index tensor $g_{\alpha\alpha^{\prime}}^s$ 
\begin{equation}
\label{stensor1}
g^s_{\alpha\alpha^{\prime}}=
\begin{cases}
a, & \alpha=\alpha^{\prime}=s=0\\
1, & \alpha=\alpha^{\prime}=s=1\\
0, & \text{otherwise}.
\end{cases}
\end{equation}
Here $a$ is a number and can be different at different positions in
order to satisfy the following condition when $K_e<0$
\begin{equation}
\label{ke_condition}
A_s|\psi\rangle=-|\psi\rangle.
\end{equation}
The action of $A_s$ is to flip four spins on the plaquette of the dual
lattice as in Fig.~\ref{ke_negative}(a).  Assuming that each
$g$-tensor at four positions $s_0,s_1,s_2,s_3$ in
Fig.~\ref{ke_negative}(a) shares the same constant $a$, then
Eq.~(\ref{ke_condition}) means that $a^{4-2n}=-1$ for
$n=0,1,\cdots,4$, where $n$ is the number of down spins in that
plaquette. However, there is no solution for the set of equations
$a^0=a^{\pm 2}=a^{\pm 4}=-1$. In conclusion, one has to break the
translation symmetry and let $a$ be different at different positions
in one plaquette. We can choose a pattern as in
Fig.~\ref{ke_negative}(b) to satisfy Eq.~(\ref{ke_condition}) for
every $s$. This choice maintains the translation invariance in the $y$
direction but doubles the unit cell in the $x$ direction; in such a
case the transfer matrix consists of two adjacent columns instead of
just one. We find that the minima of the SCL $\epsilon_{(0,k_y)}$ for
$K_e<0$ are identical with those for $K_e>0$ if $K_m$ is kept the
same. This is expected since, in our system governed by Hamiltonian
Eq.~(6) in the main text, it is the $e$-particles that appear as low
energy excitations and their dispersion is determined by the sign of
$K_m$ only (while the $m$-particles that are sensitive to the sign of
$K_e$ are still completely localized).

\begin{figure}
\begin{center}
\includegraphics[width=7.4cm]{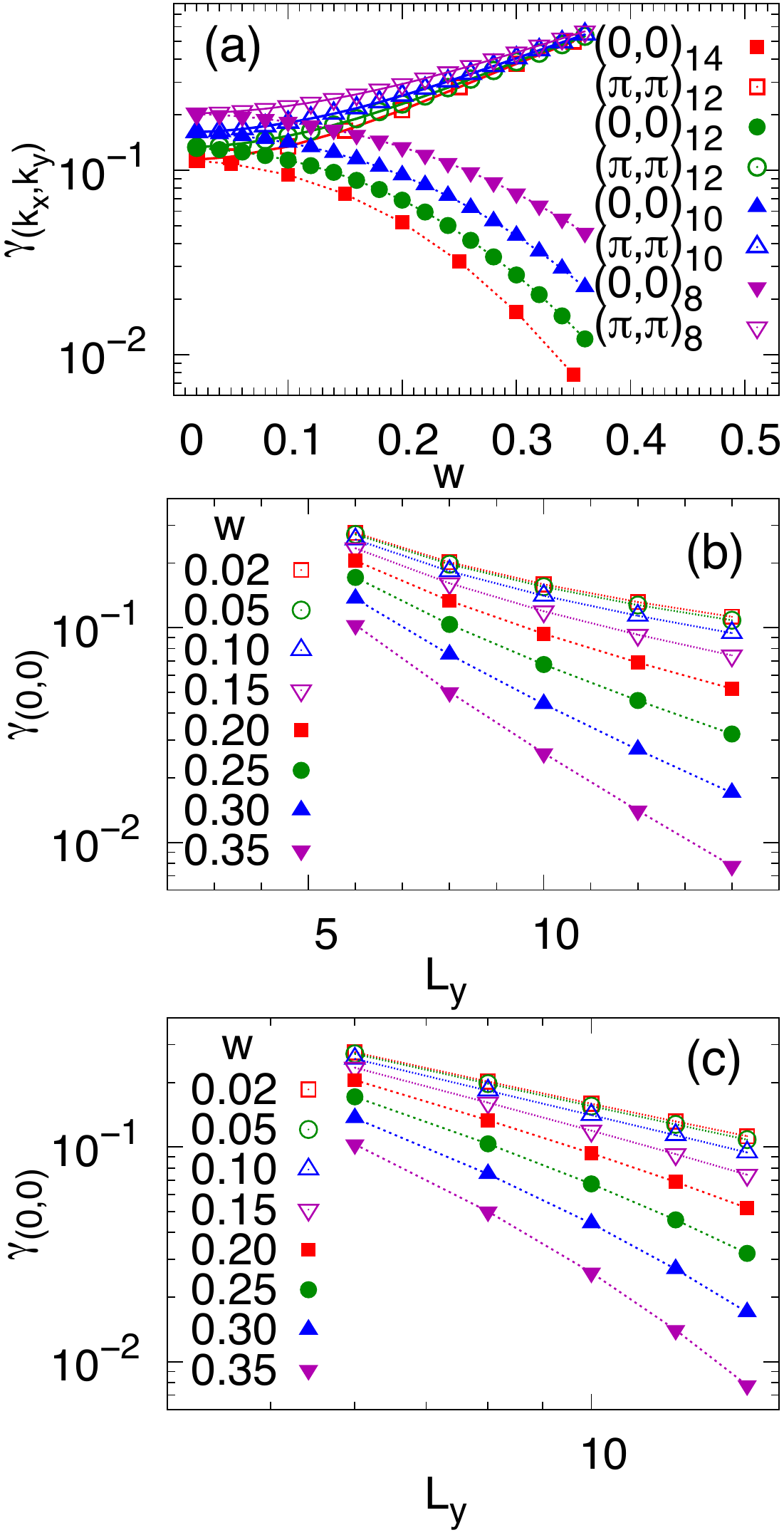}
\caption{(a) The ground state gap $\gamma_{(k_x,k_y)}$ for coupling
  $K_m<0$ $K_e>0$ as a function of $w$ at special points $(0,0)$ and
  $(\pi,\pi)$ for the cylinder perimeters $L_y=8,10,12,14$.  For
  various fixed $w$, $\gamma_{(0,0)}$ is plotted against $L_y$ in (b)
  a semi-log plot and (c) a log-log plot.}
\label{gap}
\end{center}
\vskip -0.5cm
\end{figure}

\subsection{Phase diagram of the PEPS wavefunction by tuning parameter
  $w$ for $K_m<0$ and $K_e>0$}
This section studies the phase diagram of the PEPS
wavefunction by tuning parameter $w$ [in the local operator
$\text{diag}(w,1)$] from 1 to 0 for $K_m<0$, $K_e>0$. In one limit,
when $w\approx 1$, the state is deep in the topologically ordered
phase. In the opposite limit, $w\to
0$, all spins tend to point down, but the constraint
Eq.~\ref{ttensor1} requires that each plaquette has an odd
number of up spins.  These conditions together lead to a state where all
configurations with one spin up in each plaquette contribute equally.

When $w$ is in between 0 and 1, one can imagine that the entropy
contributed from the massive number of configurations that meet
Eq.~\ref{ttensor1} is competing with the Zeeman energy, and a valence
bond solid order could emerge. As one varies $w$ from 1 to 0, the PEPS
wavefunction could go through a phase transition from the
topologically ordered phase to a valence bond solid phase. Inside the
valence bond solid phase, translation symmetry is spontaneous broken
in the thermodynamic limit, but at any finite size, the ground states
are degenerate. We expect that two degenerate eigenvalues with $\pm 1$
will appear as the leading eigenvalues of the transfer matrix; whereas
in the topological phase, both leading eigenvalues are $+1$.

We now try to identify this phase transition using techniques
discussed in the paper. We define $\gamma_{(k_x,k_y)}$, which roughly
represents the gap to the ground state at momentum $(k_x,k_y)$, using
eigenvalues of the transfer matrix:
\begin{equation}
\gamma_{(k_x,k_y)}=
\begin{cases}
-\text{ln}(|\lambda^{1}_{(k_x,k_y)}|/\lambda_0),& k_x=k_y=0\\
-\text{ln}(|\lambda^{\text{max}}_{(k_x,k_y)}|/\lambda_0),& \text{otherwise}
\end{cases},
\end{equation}
[in the sector $(0,0)$ we take the subleading eigenvalue
$\lambda^{1}_{(k_x,k_y)}$, otherwise the leading eigenvalue
$\lambda^{\text{max}}_{(k_x,k_y)}$]. $\gamma_{(k_x,k_y)}$ is different
from $\epsilon_{(k_x,k_y)}$ defined earlier in that the splitting of
 $\lambda_0$, doubly degenerate in the thermodynamic limit in the
topological phase, is explicitly calibrated by $\gamma_{(0,0)}$. We
plot $\gamma_{(k_x,k_y)}$ at special points $(0,0)$ and $(\pi,\pi)$ as
a function of $w$ for cylinder perimeters $L_y=8,10,12,14$ in
Fig.~\ref{gap}(a).  Note that all other gaps lie well above these two
within the plotted range. We do not see a crossing of
$\gamma_{(0,0)}$ and $\gamma_{(\pi,\pi)}$ as we vary $w$.  Instead, we
find that they approacheach other, with $\gamma_{(\pi,\pi)}$
slightly larger than $\gamma_{(0,0)}$ when $w$ is close to 0. In order
to examine finite size effects we plot $\gamma_{(0,0)}$ as a
function of $L_y$ at various fixed $w$, in a semi-log plot [in
Fig.~\ref{gap}(b)] and a log-log plot [in Fig.~\ref{gap}(c)]. The results are inconclusive; we cannot identify any critical
point where a topological phase turns into a valence bond solid phase
due to very large correlation length and insufficient system $L_y$
size available.

\subsection{Diagonalizing the transfer matrix of a cylinder in
  momentum basis}
In the general case, the transfer matrix, denoted as $\mathbb{E}$, of
a cylinder of circumference $L_y$ can be thought of as a Hamiltonian
for a system composed of local degrees of freedom with Hilbert space
dimension $D^2$ ($D$ is the bond dimension of PEPS) where each degree
of freedom is composed of two virtual degrees of freedom.  These local
degrees of freedom are arranged in a ring with $L_y$ ``sites''. Note
that the transfer matrix is generally not hermitian. The matrix
elements of the transfer matrix in real space can be computed by
specifying a basis vector $|a\rangle$ (which denotes compactly
configuration of all $L_y$ local degrees of freedom) and multiplying
it by the transfer matrix
\begin{equation}
  \mathbb{E}|a\rangle=\sum_bh_{ba}|b\rangle,
\end{equation}
where $h_{ba}=\langle b|\mathbb{E}|a\rangle$ is the matrix
element. 

The transfer matrix commutes with translations $T$ in the $y$
direction and hence can be diagonalized simultaneously with $T$.  The
eigenvalues of $T$ have the form $e^{ik}$ where $k = 2\pi m/L_y$, $m =
0, 1, \dots, L_y-1$.  We construct the corresponding eigenspace $V_k$
as follows.

First, we consider classes of real-space configurations that are
related to each other by an action of $T$: $|a\rangle$ and
$|a'\rangle$ belong to the same class if $|a'\rangle=\ T^\ell
|a\rangle$ for some $\ell$.  It is easy to see that we can specify
each class by one representative member $|a\rangle$, while all other
members are given by $T^\ell|a\rangle$, $\ell=0,1, \dots R_a-1$, where
$R_a$ is the smallest number such that $T^{R_{a}}|a\rangle =
|a\rangle$; the ``periodicity'' $R_a$ must be less than or equal to
$L_y$ since $T^{L_y}=1$.  For a given momentum $k$, if $k R_a$ is an
integer multiple of $2\pi$, then out of the states in this class we
can construct a single normalized eigenstate of $T$ with eigenvalue
$e^{ik}$ as follows:
\begin{equation}
\label{rmatrixelement}
|a_k\rangle = \frac{1}{\sqrt{N_a}}\sum_{r=0}^{L_y-1}e^{-ikr}T^r|a\rangle,
\end{equation}
where $N_a$ is the normalization constant
\begin{equation}
N_a = L_y^2/R_a.
\end{equation}
On the other hand, if $k R_a$ is not an integer multiple of $2\pi$, we
cannot construct such an eigenstate from the states in this class;
thus this class of configurations is not compatible with the momentum
$k$ and is not included as a basis state for $V_k$.  By going over all
classes that are compatible with $k$, we construct a complete basis of
$V_k$, which we call the momentum basis.

We calculate the transfer matrix elements in the basis of $V_k$
as follows
\begin{eqnarray}
\nonumber
\mathbb{E}|a_k\rangle&=&\frac{1}{\sqrt{N_a}}\sum_{r=0}^{L_y-1}e^{-ikr}T^r\mathbb{E}|a\rangle\\
\label{kmatrixelement}
&=&\sum_{b^{\prime}}h_{b^{\prime}a}\frac{1}{\sqrt{N_a}}\sum_{r=0}^{L_y-1}e^{-ikr}T^r|b^{\prime}\rangle,
\end{eqnarray}
where $|b^{\prime}\rangle$ runs over all real-space basis states.
Each such $|b^{\prime}\rangle$ belongs to some class whose
representative we denote as $|b\rangle$, and hence there is
$l_{b^\prime}$ such that
\begin{equation}
|b^{\prime}\rangle = T^{-l_{b^{\prime}}} |b\rangle.
\end{equation}
Substituting this in Eq.~(\ref{kmatrixelement}), we have
\begin{equation}
\mathbb{E}|a_k\rangle=\sum_{b^{\prime}}h_{b^{\prime}a}e^{-ikl_{b^{\prime}}}\frac{\sqrt{N_b}}{\sqrt{N_a}}|b_k\rangle,
\end{equation}
where the summation runs over all real-space basis states
$|b^{\prime}\rangle$ and not only the representative $b\rangle$.  Note
that for each momentum-space basis state $|b_k \rangle$, the matrix
element $\langle a_k| \mathbb{E} |b_k\rangle$ obtains contributions
from all $b'$ that belong to the class which corresponds to $|b_k
\rangle$.

After finding the matrix elements $\langle b_k|\mathbb{E}|a_k\rangle$
in the basis in $V_k$, we can separately diagonalize the block
diagonal transfer matrix for each momentum $k$ and obtain the
corresponding $\epsilon_{k_y}$.

In the present case, the transfer matrix actually simplifies since the
structure of the $T$ and $g$ tensors are such that the labels
$\alpha\alpha^{\prime}$ of the double tensor must coincide at each
``site'' in the transfer matrix; $\alpha_i=\alpha^{\prime}_i$ and
$\beta_i=\beta^{\prime}_i$ ($i=0,1,\cdots,L_y-1$), thus the size of
Hilbert space is reduced from $D^{2L_y}$ to $D^{L_y}$.

\end{document}